\documentclass[]{tLCT2e}

\usepackage{amsmath,amssymb,amsthm}
\usepackage{mathrsfs}
\usepackage{hyperref}
\usepackage{graphicx}
\usepackage{subcaption}
\usepackage{mdwlist}
\usepackage{bm}
\usepackage{empheq}
\usepackage{verbatim}
\usepackage{paralist}

\usepackage[usenames]{color}

\newtheorem{thm}{Theorem}
\newtheorem{prop}[thm]{Proposition}

\newcommand{\das}{:=}
\newcommand{\T}{^\mathsf{T}} 
\newcommand{\N}{\mathbb{N}}

\newcommand{\R}{\mathbb{R}}

\renewcommand{\leq}{\leqslant}
\renewcommand{\geq}{\geqslant}

\newcommand{\grad}{\nabla}               
\newcommand{\dv}{\operatorname{div}}     
\newcommand{\curl}{\operatorname{curl}}  
\newcommand{\tr}{\operatorname{tr}}      

\DeclareMathOperator{\epr}{\times} 

\newcommand{\body}{\mathscr{B}}
\newcommand{\boundary}{\partial\body}
\newcommand{\area}{A}
\newcommand{\volume}{V}
\newcommand{\n}{\bm{n}}
\newcommand{\normal}{\bm{\nu}}

\newcommand{\e}{\bm{e}}
\newcommand{\alin}{a_\mathrm{lin}}
\newcommand{\abar}{a_\mathrm{qua}}
\newcommand{\grads}{\grad_{\!\mathrm{s}}}
\newcommand{\dd}{\operatorname{d}\!}
\newcommand{\dvs}{\dv_{\!\mathrm{s}}}
\newcommand{\curls}{\curl_{\mathrm{s}}}
\newcommand{\free}{\mathcal{F}}
\newcommand{\Free}{\mathscr{F}}
\newcommand{\gradnu}{\grad_{\!\bm{\nu}}}
\newcommand{\dvnu}{\dv_{\!\bm{\nu}}}
\newcommand{\id}{\mathbf{I}}
\newcommand{\torframe}{(\e_r,\e_\varphi,\e_\psi)}
\newcommand{\torcoo}{(r,\varphi,\psi)}
\newcommand{\carframe}{(\e_x,\e_y,\e_z)}

\begin{document}
	\title{Instability of toroidal nematics}
\author{Andrea Pedrini and Epifanio G. Virga$^{\ast}$\thanks{$^\ast$Corresponding author. Email: eg.virga@unipv.it
		\vspace{6pt}} \\
		\vspace{6pt}
		{\em{Dipartimento di Matematica, Universit\`a di Pavia, via Ferrata 5, 27100 Pavia, Italy}}\\
		\received{Received 00 Month 20XX; final version received 00 Month 20XX} }
	\date{\today}
	\maketitle
\begin{abstract}
Toroidal nematics are nematic liquid crystals confined within a circular torus and subject to planar degenerate anchoring on the boundary of the torus. They may be droplets floating in an isotropic environment or cavities carved out of a solid substrate. A universal solution of Frank's elastic free energy is an equilibrium configuration for the nematic director field, irrespective of the values of the elastic constants, whose vector lines are the coaxial parallels of the torus. We explore the local stability of this configuration and identify a range of parameters where the main drive towards instability does not come from  the surface-like elastic constant $K_{24}$ being large, but from the the ratio $K_2/K_3$ of the twist to bend elastic constants being small, which also  makes our study relevant to chromonic liquid crystals.
\end{abstract}

	



\section{Introduction}
 \label{sec:intro}


Nematic liquid crystals are characterized by a natural, undistorted state, where the director $\n$ is uniform in space (in an arbitrarily chosen direction). The most elementary measure of distortion is therefore the spatial gradient $\grad\n$. The simplest formula for the elastic free energy density (per unit volume) was put forward by Frank~\cite{frank:theory}: it is the most general quadratic expression in $\grad\n$ invariant under rotations and complying with the \emph{nematic} symmetry, embodied by the director reversion, $\n\mapsto-\n$. Frank's free-energy functional $\Free$ is given by   
\begin{equation}\label{eq:Frank_free_energy}
\begin{split}
\Free[\bm{n}] = \int_{\body}\Big\{&
\frac{1}{2}\left[K_{1}(\dv\bm{n})^{2} + K_{2}(\bm{n}\cdot\curl\bm{n})^{2} + K_{3}|\bm{n}\epr\curl\bm{n}|^{2}\right]\\&+
K_{24}[\tr(\grad\bm{n})^{2}-(\dv\bm{n})^{2}]\Big\}\dd\volume\,,
\end{split}
\end{equation}
where $\body$ is the region in space occupied by the material and $\dd\volume$ denotes the volume element. Classically, to render \eqref{eq:Frank_free_energy} more symmetric, a constant $K_4$ is often introduced so that $K_{24}=\frac12(K_2+K_4)$. 

$K_1$, $K_2$, and $K_3$ are often referred to as the \emph{splay}, \emph{twist}, and \emph{bend} elastic constants, respectively, by the names given to the three fundamental distortion modes characterized by the excitation of only the corresponding energy (see, for example, \S\,2.2 of \cite{stewart:static} or \S\,3.3 of \cite{virga:variational} for both a derivation of \eqref{eq:Frank_free_energy} and the traditional pictorial description of the fundamental distortion modes).

The elastic constant $K_{24}$ is notoriously different from the others in at least two respects,
\begin{inparaenum}[(i)]
\item it weights an energy that can only be approximately isolated by the \emph{saddle-splay} distortion \cite[p.\,121]{virga:variational} and
\item it can be converted into a surface integral over the boundary $\boundary$ of the domain occupied by the material.
\end{inparaenum}
The latter property says that the saddle-splay energy is a \emph{null Lagrangian}, which does not contribute to the Euler-Lagrange equation obeyed by the equilibrium nematic textures \cite{ericksen:nilpotent}. Moreover, it can be shown that the $K_{24}$-energy, once converted into a surface integral, depends only on $\n$ and its surface gradient $\grads\n$ on $\boundary$. Thus, we finally give $\Free$ in \eqref{eq:Frank_free_energy} the following form,
\begin{equation}\label{eq:frank}
\begin{split}
 \Free[\bm{n}] = &\frac{1}{2} \int_{\body}\left[[K_{1}(\dv\bm{n})^{2} + K_{2}(\bm{n}\cdot\curl\bm{n})^{2} + K_{3}|(\grad\bm{n})\bm{n}|^{2}\right]\dd\volume \\
 &+  \int_{\boundary}K_{24}[(\grads\bm{n})\bm{n}-(\dvs\bm{n})\bm{n}]\cdot\bm{\nu}\dd\area\,,
\end{split}
\end{equation}
where $\dd\area$ is the area element and $\bm{\nu}$ is the outer unit normal to $\boundary$.

Ericksen~\cite{ericksen:inequalities} first remarked that the free-energy density associated with $\Free$ in \eqref{eq:Frank_free_energy} is positive semi-definite, duly representing the cost incurred in distorting the natural uniform state, only if the elastic constants obey the inequalities 
\begin{equation}\label{eq:Ericksen_inequalities}
K_{1}\geq K_{24}\geq0,\quad K_{2}\geq K_{24},\quad K_{3}\geq 0\,,
\end{equation}
which are also referred to as Ericksen's inequalities.

Two different types of boundary conditions for $\n$ on $\boundary$ give the $K_{24}$-energy  a special form: these are the \emph{strong} and the \emph{planar degenerate} anchorings. In the former case, $\n$ is prescribed on the whole of $\boundary$, and so the $K_{24}$-integral is the same for all competing equilibrium textures, and it can be altogether ignored. In the latter case, 
\begin{equation}\label{eq:planar_degenerate}
\n\cdot\normal\equiv0\quad\text{on}\quad\boundary
\end{equation}
and, as remarked in \cite{koning:saddle-splay}, the $K_{24}$-integral can be rewritten as 
\begin{equation}\label{eq:K_24_geometric}
-K_{24}\int_{\boundary}\left(\kappa_1n_1^2+\kappa_2n_2^2\right)\dd\area\,,
\end{equation}
where $\kappa_1$ and $\kappa_2$ are the principal curvatures of $\boundary$, and $n_i$ are the components of $\n$ along the corresponding principal directions of curvature.\footnote{We write the curvature tensor as $\grads\normal=\kappa_1\e_1\otimes\e_1+\kappa_2\e_2\otimes\e_2$, where $\e_1$ and $\e_2$ are unit vectors along the principal directions of curvature of $\boundary$.} It is clear from \eqref{eq:K_24_geometric} that for $K_{24}>0$, which is the strong form of  \eqref{eq:Ericksen_inequalities}, whenever  \eqref{eq:planar_degenerate} applies the saddle-splay energy would locally tend to orient $\n$ on $\boundary$ along the direction of \emph{maximum} (signed) curvature. We shall see below the form that \eqref{eq:K_24_geometric} takes in the stability problem studied here.

To close our hasty introduction to the fundamentals of the mathematical theory of nematic liquid crystals and to place our study in a broader perspective, we recall the meaning of \emph{universal solutions} in the hydrostatics of liquid crystals. They  were first considered by Ericksen~\cite{ericksen:general} for a general elastic free-energy density delivered by an isotropic function $W=W(\n,\grad\n)$, not necessarily quadratic in $\grad\n$. Ericksen proved that a locally smooth director field $\n$ solving the Euler-Lagrange equation associated with all such functions may only have \emph{rectilinear} vector lines, which are either
\begin{enumerate}[(i)]
	\item parallel straight lines,
	\item straight lines precessing in a pure twist with director cosines $(\cos\mu z,\sin\mu z,0)$, where $\mu$ is a constant,
	\item lines orthogonal to a family of concentric spheres or
	\item lines orthogonal to a family of coaxial cylinders.
	\suspend{enumerate}
	Marris~\cite{marris:universal,marris:addition}  proved that the extra universal solutions afforded by Frank's expression for the elastic free-energy density, which then solve the Euler-Lagrange equation for $\Free$ in \eqref{eq:Frank_free_energy} for all values of Frank's constants, are fields obtained by translating uniformly in space planar fields whose vector lines are either
	\resume{enumerate}[{[(i)]}]
	\item \label{item:Marris_1} concentric circles or
	\item \label{item:Marris_2} coaxial circles intersected orthogonally by all members of another family of coaxial circles. 
\end{enumerate}
Family (\ref{item:Marris_1}) comprises a configuration that shall particularly interest us here. Family (\ref{item:Marris_2}) exerts quite a classic fascination, as both pencils of circles featuring as vector lines of $\n$ are Apollonian circles (see, for example, \S\,2 of \cite{ogilvy:excursions}).

Despite their stance in the mathematical theory of liquid crystals, little is known about the stability of universal solutions. This paper is concerned with the local stability analysis of universal solution \eqref{item:Marris_1} confined within a torus with planar degenerate anchoring on its boundary. 
This is precisely what we mean by \emph{toroidal nematics}. They are theoretical constructs, which in reality may be equally approximated by either toroidal droplets produced in an isotropic environment or toroidal cavities carved out of a rigid substrate, both  inducing a degenerate planar anchoring (with no preferred surface orientation).

We build on a previous pioneering study \cite{koning:saddle-splay}, from which we mainly drew our inspiration, and which we think have contributed to improve. The merit of \cite{koning:saddle-splay} is having shown the role of $K_{24}$ as a driving force behind the destabilization of a universal solution. However, the method through which such a role is explored is too special to be credit with a universal meaning; the second variation of $\Free$ is simply shown to become negative for a cleverly chosen, but particular test function. We were similarly unable to characterize the sign of the second variation of $\Free$, which would have afforded a complete local stability analysis, but we have succeeded in extending significantly the pool of destabilizing modes, thus refining the physical interpretation of the destabilizing causes. Our instability criterion systematically improves on that found in \cite{koning:saddle-splay}.

The paper has the following structure. Section~\ref{sec:toroidal}, which is preparatory in nature, is concerned with the representation of nematic director fields in the unusual toroidal geometry. In Sect.~\ref{sec:variation}, we compute in general the second variation of $\Free$ on the universal solution \eqref{item:Marris_1} and, following a fortunate intuition of \cite{koning:saddle-splay}, we specialize it to divergence-free test functions. In Sect.~\ref{sec:criteria}, we benefit from this representation of the second variation and compute it for both the test function used in \cite{koning:saddle-splay} and a new class of test functions, which eventually afford an improvement on their instability criterion. Finally, in Sect.~\ref{sec:conclusion}, we collect the main conclusions of our study and try and extract suggestions from the new perspective gained here on how to attack the true challenge that still faces us: the non-linear stability of toroidal nematics and the minimizers they fall in, once their universal solution has been destabilized. A technical appendix closes the paper, where all tedious, but necessary mathematical details are relegated for the perusal of the interested reader who wishes to follow all our steps.

\section{Toroidal nematic fields}
 \label{sec:toroidal}
This section is mainly descriptive in nature; it is concerned with the construction of a family of nematic director fields $\n$ inspired by the toroidal symmetry of the domain $\body$ where the free energy $\Free$ will be studied.
\subsection{Toroidal coordinates and frame}\label{sec:toroidal_coordinaites}
\begin{figure}[h!]
\centering
\includegraphics[width=.7\linewidth]{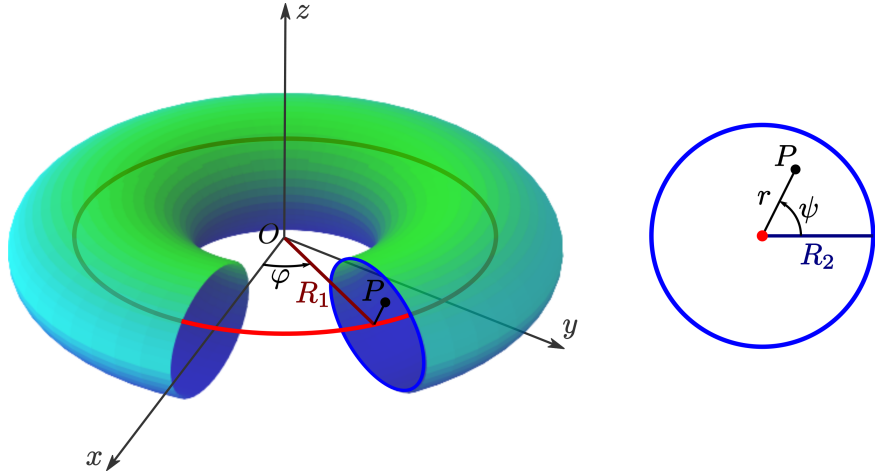}
\caption{The geometric representation of the surface of a circular torus $\body$, to be identified with either a droplet or a cavity, obtained by revolving a circle of radius $R_{2}$ around the $z$-axis. The position of a point $P$ in $\body$ can be parameterized by $r$, $\varphi$ and $\psi$, as in~\eqref{eq:positionvector}. On the left: geometric representation of the entire surface and the center line with radius $R_{1}$ (in red), with azimuthal angle $\varphi$. On the right: geometric representation of a cross-section with radius $R_{2}$ of the toroidal surface (in blue), with radial coordinate $r$ and polar angle $\psi$.}
\label{fig:toruscoordinates}
\end{figure}

Figure~\ref{fig:toruscoordinates} depicts a circular torus $\body$, which is to be thought of as either a droplet or a cavity whose boundary enforces the \emph{degenerate planar} boundary condition on $\n$, requiring only that $\n\cdot\normal\equiv0$, while leaving $\n$ otherwise unspecified.
In Fig.~\ref{fig:toruscoordinates}, $O$ is the center of $\body$
with minor and major radii $R_{2}>0$ and $R_{1}>R_2$, respectively. $\body$ is axially symmetric about the $z$-axis of a standard (orthonormal and positively oriented) Cartesian frame $\carframe$ in three-dimensional space. The position vector $\bm{p}$ of a generic point $P$ within $\body$ issued from  the origin $O$ can be expressed as
\begin{equation}\label{eq:positionvector}
\bm{p} \das P-O = (R_{1}+r\cos\psi)\cos\varphi\,\bm{e}_{x} + (R_{1}+r\cos\psi)\sin\varphi\,\bm{e}_{y} + r\sin\psi\,\bm{e}_{z}\,,
\end{equation}
where $r\in[0,R_{2}]$ is the \emph{radial coordinate}, $\varphi\in[0,2\pi)$ is the \emph{azimuthal angle} and $\psi\in[0,2\pi)$ is the \emph{polar angle}. 

\begin{figure}[h!]
  \centering
  \includegraphics[width=.7\linewidth]{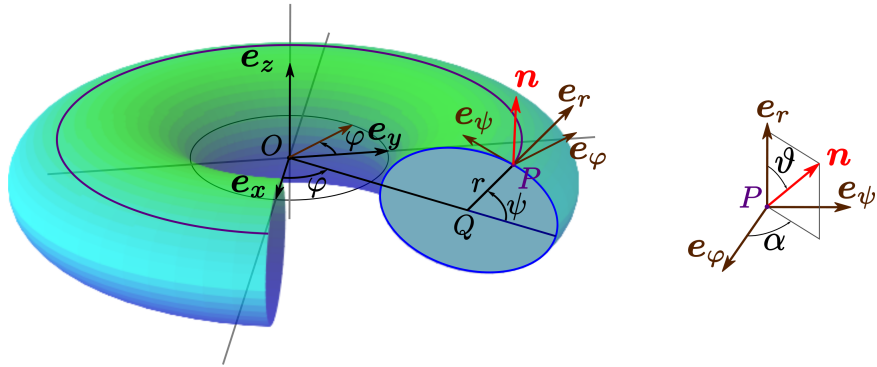}
  \caption{On the left, the toroidal frame $(\bm{e}_{r},\bm{e}_{\varphi},\bm{e}_{\psi})$ at a point $P$ within the torus $\body$: $\bm{e}_{r}$ has the same direction of the vector $P-Q$,  $\bm{e}_{\varphi}$ is orthogonal to the plane containing both the $z$ axis and $P$, and $\bm{e}_{\psi}$ is tangent to the disk of radius $r$ and centre $Q$ on that very plane. On the right, the angles $\alpha$ and $\vartheta$ used to express the director field $\bm{n}$ in the toroidal frame.}
\label{fig:toruslocal}
\end{figure}

At any point $P$ in $\body$,  we  introduce the (orthonormal and positively oriented) \emph{toroidal frame} $(\bm{e}_{r},\bm{e}_{\varphi},\bm{e}_{\psi})$ conjugated with the \emph{toroidal coordinates} $(r,\varphi,\psi)$. As  shown in Fig.~\ref{fig:toruslocal}, the toroidal frame can be expressed in the Cartesian frame $\carframe$ through the equations
\begin{equation}
\label{eq:localbasis}
 \left\{\!
 \begin{aligned}
  \bm{e}_{r} &\das \cos\varphi\cos\psi\,\bm{e}_{x}+\sin\varphi\cos\psi\,\bm{e}_{y}+\sin\psi\,\bm{e}_{z}\,,\\
  \bm{e}_{\varphi} &\das -\sin\varphi\,\bm{e}_{x}+\cos\varphi\,\bm{e}_{y}\,,\\
  \bm{e}_{\psi} &\das -\cos\varphi\sin\psi\,\bm{e}_{x}-\sin\varphi\sin\psi\,\bm{e}_{y}+\cos\psi\,\bm{e}_{z}\,,
 \end{aligned}
 \right.
\end{equation}
see Appendix~\ref{app:geometrylocalbasis} for the analytic details that Fig.~\ref{fig:toruslocal} cannot convey.
In  toroidal coordinates,  the volume and area elements are
\begin{equation}\label{eq:area_volume_elements}
\dd\volume = r(R_{1}+r\cos\psi)\dd r\dd\varphi \dd\psi \quad\text{and}\quad \dd\area = r(R_{1}+r\cos\psi)\dd\varphi \dd\psi\,.
\end{equation}

In the toroidal frame, the director field can be expressed as 
\begin{equation}
\label{eq:director}
 \bm{n} =n_{r}\,\bm{e}_{r} +n_{\varphi}\,\bm{e}_{\varphi} + n_{\psi}\,\bm{e}_{\psi}\,. 
 \end{equation}
In particular,  taking $\alpha$ and $\vartheta$ to be the angles defined as in Fig.~\ref{fig:toruslocal}, we can write
\begin{equation}\label{eq:angular_representation}
n_{r} = \cos\vartheta,\quad n_{\varphi} = \cos\alpha\sin\vartheta,\quad \text{and}\quad n_{\psi} = \sin\alpha\sin\vartheta\,.
\end{equation}
As complicated as the problem of minimizing  $\Free$ in the torus $\body$ may be, we know at least that the director field $\n$ in \eqref{eq:director} with $n_r=n_\psi\equiv0$ is an equilibrium configuration for $\Free$, which may or may not be a minimizer. As recalled in the Introduction,  $\n\equiv\e_\varphi$ is a universal solution of $\Free$, which here we shall call the \emph{axial configuration}, for brevity.

\subsection{Toroidal fields}\label{sec:toroidal_fields}
Here the stability of the axial configuration will be probed within a large, but restricted class of director fields which share one essential feature with the perturbed field: they are everywhere tangent to either the torus $\body$ or any of its inwards. The axial configuration is generated by setting $\alpha\equiv0$ and $\vartheta\equiv\frac\pi2$: its vector lines are the \emph{parallels} of the torus, an  \emph{achiral} pattern, invariant  under central inversion about $O$. An intrinsic way to measure the \emph{chirality} of the vector lines of  $\n$ is computing their \emph{helicity}\footnote{We learn from Truesdell~\cite[p.\,332]{truesdell:first} that this definition of $\Omega$   was introduced by Zhukowsky and named \emph{abnormality} by Levi-Civita. We shall use the modern name of \emph{helicity}.} $\Omega$,  
\begin{equation}\label{eq:abnormality}
\Omega:=\n\cdot\curl\n\,.
\end{equation}
For the axial configuration, clearly $\Omega=0$. 

As in \cite{koning:saddle-splay}, we shall also take the  director field to be independent of $\varphi$, subject to the condition $\n=\e_\varphi$ for $r=0$, and with no radial component (i.e.,\ $\vartheta\equiv\frac{\pi}{2}$). Therefore
\begin{equation}\label{eq:director_toroidal_field}
\bm{n} = \cos\alpha\,\bm{e}_{\varphi} + \sin\alpha\,\bm{e}_{\psi}\,,\quad\text{with}\quad\alpha=\alpha(r,\psi)\quad\text{and}\quad\alpha(0,\psi)\equiv0\,.
\end{equation}
Specializing \eqref{eq:appendix_abnormality_n} of Appendix~\ref{app:geometrydirectorfield} to the field in \eqref{eq:director_toroidal_field}, we easily see that the vector lines of the latter possesses the helicity\footnote{Here and in the following, $f_{,x}$ denotes the partial derivative of a function $f$ with respect to one of its variables, $x$.}
\begin{equation}
\Omega=-\alpha_{,r}-\frac{R_1}{r(R_1+r\cos\psi)}\sin\alpha\cos\alpha\,.
\end{equation} 
The latter formula makes it clear that reversing the sign of $\alpha$ also reverses the sign of $\Omega$ thus producing a director field $\n$ with opposite chirality. Correspondingly, the vector lines of two conjugated director fields as in \eqref{eq:director_toroidal_field} that only differ by the sign of $\alpha$ are mapped into one another by a central inversion about $O$.

Let
\begin{equation}\label{eq:eta_definition}
\eta\das\frac{R_{2}}{R_{1}}\in[0,1]
\end{equation}
be the ratio between the radii of the torus. By performing the change of coordinate
\begin{equation}\label{eq:sigma_definition}
\sigma\das\frac{r}{R_{2}} = \frac{r}{R_{1}\eta},
\end{equation}
the angle $\alpha$ becomes a function of $\sigma$, and $\psi$ and, as shown with more details in Appendix~\ref{app:geometryfrankenergy}, Frank's elastic free-energy functional takes the following reduced form, 
\begin{equation}\label{eq:frankalpha}
\begin{split}
 \frac{\mathcal{F}[\alpha]}{\pi R_{1}} &= \int_{0}^{1}\!\int_{0}^{2\pi}\!\left\{K_{1}\frac{\left[\eta\sigma\sin\psi\sin\alpha - (1+\eta\sigma\cos\psi)\cos\alpha\,\alpha_{,\psi}\right]^{2}}{\sigma(1+\eta\sigma\cos\psi)}\right. \\
 &\qquad\qquad\qquad + K_{2} \frac{\left[\sigma(1+\eta\sigma\cos\psi)\alpha_{,\sigma} + \sin\alpha\cos\alpha\right]^{2}}{\sigma(1+\eta\sigma\cos\psi)} \\
&\qquad\qquad\qquad + K_{3} \frac{\left[\eta\sigma\sin\psi\cos\alpha + (1+\eta\sigma\cos\psi)\sin\alpha\,\alpha_{,\psi}\right]^{2}}{\sigma(1+\eta\sigma\cos\psi)} \\
 &\qquad\qquad\qquad+ \left.K_{3}\frac{\left[\eta\sigma\cos\psi + \sin^{2}\alpha\right]^{2}}{\sigma(1+\eta\sigma\cos\psi)}\right\}\dd\sigma d\psi\,, \\
 &\qquad - 2K_{24} \int_{0}^{2\pi}\sin^{2}\alpha(1,\psi)\dd\psi\,,
\end{split}
\end{equation}
where the last integral reflects the general surface energy \eqref{eq:K_24_geometric} and its ability to destabilize the axial configuration for $K_{24}>0$.

It was also remarked in \cite{koning:saddle-splay} that the choice of the field in \eqref{eq:director_toroidal_field} can be further simplified by seeking guidance in the explicit form \eqref{eq:frankalpha} of the functional $\free$. By requiring $\n$ to be divergence-free, not only we keep the perturbing director field in the same family as the axial configuration which we intend to perturb, but we also spare an energetic contribution, the one coming from the splay constant $K_1$, which may be dominant, as is, for example, the case for the newly discovered chromonic liquid crystals, for which typically $K_1\approx K_3\approx10K_2$ \cite{zhou:elasticity}.\footnote{We were not able to prove that a divergence-free field $\n$ represents the optimal way to probe the stability of the axial configuration. At this stage, the example of chromonics is only meant to be suggestive. They indicate that nematic liquid crystals with high elastic anisotropies are not just a theoretical curiosity, but real-life materials.} By requiring that $\dv\n=0$, also with the aid of \eqref{eq:appendix_div_n}, we obtain the following differential equation
\begin{equation}\label{eq:divergence_free_condition}
\alpha_{,\psi} = \frac{\eta\sigma\sin\psi\tan\alpha}{1+\eta\sigma\cos\psi}\,,
\end{equation}
which can be easily integrated, delivering
\begin{equation}\label{eq:divergence_free_alpha}
 \alpha=\arcsin\frac{a(\sigma)}{1+\eta\sigma\cos\psi}\,,
\end{equation}
where $a(\sigma)$ is a real function of $\sigma$ only. For \eqref{eq:divergence_free_alpha} to obey the inequality
$|\sin\alpha|\leq1$, $a$ must be such that  $|a(\sigma)|\leq 1+\eta\sigma\cos\psi$ for all $\psi\in[0,2\pi]$, whence it follows that  
\begin{equation}
\label{eq:barrierscond}
 |a(\sigma)|\leq1-\eta\sigma\,.
\end{equation}

In the following section, we shall make use of \eqref{eq:divergence_free_alpha} in \eqref{eq:frankalpha} and we shall write $\free$ in the quadratic approximation for $\alpha$, so as to convert it into the second variation of the elastic free-energy evaluated at the axial configuration.

\section{Second free-energy variation}\label{sec:variation}
For $\alpha\equiv0$, $\free[0]$ delivers the energy associated with the axial solution, which has pure bend (see also Appendix~\ref{app:geometryfrankenergy}),
\begin{equation}\label{eq:pure_bend_energy}
\free[0]=2\pi^2R_1K_3(1-\sqrt{1-\eta^2})\,.
\end{equation}
The extra elastic energy associated with a distortion other than $\alpha\equiv0$ will be denoted by
\begin{equation}\label{eq:extra_energy}
\mathcal{E}[\alpha]:=\free[\alpha]-\free[0]\,.
\end{equation}
Taking the quadratic approximation for $\alpha$ in \eqref{eq:frankalpha}, we effectively compute the second variation of $\mathcal{E}$ at $\alpha\equiv0$,
\begin{equation}
\label{eq:franktaylor}
\begin{split}
 \frac{\mathcal{E}[\alpha]}{\pi R_{1}} &= \int_{0}^{1}\!\int_{0}^{2\pi}\left\{K_{1}\left[\frac{\sqrt{\eta\sigma}\sin\psi}{\sqrt{1+\eta\sigma\cos\psi}}\alpha - \frac{\sqrt{1+\eta\sigma\cos\psi}}{\sqrt{\eta\sigma}}\alpha_{,\psi}\right]^2\right.\\
 &\qquad\qquad\qquad+ K_{2} \left[\frac{1}{\sqrt{\eta\sigma}\sqrt{1+\eta\sigma\cos\psi}}\alpha + \frac{\sqrt{\eta\sigma}\sqrt{1+\eta\sigma\cos\psi}}{\eta}\alpha_{,\sigma}\right]^2\\
 &\qquad\qquad\qquad+ \left.K_{3} \frac{\cos\psi - \eta\sigma}{1+\eta\sigma\cos\psi}\alpha^{2} \right\}\eta\dd\sigma d\psi\\
&\qquad- 2K_{24}\int_{0}^{2\pi}\alpha^2(1,\psi)\dd\psi
\end{split}
\end{equation}
(see Appendix~\ref{app:geometrytaylor} for more details). Enforcing in this context the divergence-free condition for $\n$ amounts to linearize \eqref{eq:divergence_free_alpha} in $a$, which thus becomes
\begin{equation}\label{eq:divergence_free_linearized}
 \alpha=\frac{a(\sigma)}{1+\eta\sigma\cos\psi}\,,
\end{equation}
which is still subject to \eqref{eq:barrierscond}. For this latter to be valid up to $\sigma=1$ and $a(1)$ to remain free, though infinitesimal, we shall hereafter take $0\leq\eta<1$. Making use of \eqref{eq:divergence_free_linearized} in \eqref{eq:franktaylor}, we readily arrive at
\begin{equation}
\label{eq:energysigma}
\begin{aligned}
 \frac{\mathcal{E}[a]}{\pi R_{1}} &= \int_{0}^{1}\!\int_{0}^{2\pi}\left\{K_{2} \left[\frac{1-\eta\sigma\cos\psi}{\sqrt{\eta\sigma}(\sqrt{1+\eta\sigma\cos\psi})^{3}}a(\sigma) + \frac{\sqrt{\eta\sigma}}{\eta\sqrt{1+\eta\sigma\cos\psi}}a_{,\sigma}(\sigma)\right]^2\right.\\
 &\qquad\qquad\qquad\left.+ K_{3} \frac{\cos\psi - \eta\sigma}{(1+\eta\sigma\cos\psi)^{3}}a^{2}(\sigma) \right\}\eta \dd\sigma \dd\psi\\
&\qquad- 2K_{24}\int_{0}^{2\pi}\frac{1}{(1+\eta\cos\psi)^{2}}a^2(1) \dd\psi\,.
\end{aligned}
\end{equation}
This form of $\mathcal{E}$ reveals that $a$ must vanish as $\sigma\to0$ for the twist energy to be finite, and so we shall hereafter assume that 
\begin{equation}\label{eq:a_0=0}
a(0)=0,
\end{equation} 
as a requirement of convergence for $\mathcal{E}$. A number of manipulations, which also rely on \eqref{eq:a_0=0}, are needed to convert \eqref{eq:energysigma} into a  functional for the classical Bolza problem of the Calculus of Variations in one dimension. All details are given in Appendix~\ref{app:geometrytaylor}, here we record only the final form of $\mathcal{E}$:
\begin{equation}\label{eq:energyintegrated}
\begin{split}
 \frac{\mathcal{E}[a]}{2\pi^{2}R_{1} K_{2}} &= \int_{0}^{1}\left\{ \frac{2(1+\eta^{2}\sigma^{2})^{2}+4\eta^{2}\sigma^{2}-\eta^{2}\sigma^{2}\left(2+k_{3}\right)(5+\eta^{2}\sigma^{2})}{2\sigma(1-\eta^{2}\sigma^{2})^{\frac{5}{2}}}a^{2}(\sigma)\right.\\
 &\qquad\qquad +\left. \frac{\sigma}{(1-\eta^{2}\sigma^{2})^{\frac{1}{2}}}a^{2}_{,\sigma}(\sigma) \right\} \dd\sigma + \frac{1+\eta^{2}-2k_{24}}{(1-\eta^{2})^{\frac{3}{2}}}a^{2}(1)\,,
 \end{split}
\end{equation}
where we have scaled the elastic constants to $K_2$, assumed to be positive,
\begin{equation}\label{eq:scaled_elastic_constants}
k_3:=\frac{K_3}{K_2}\quad\text{and}\quad k_{24}:=\frac{K_{24}}{K_2},
\end{equation}
so that by \eqref{eq:Ericksen_inequalities}
\begin{equation}\label{eq:rescaled_Ericksen_inequalities}
k_3\geqslant0\quad\text{and}\quad 0\leqslant k_{24}\leqslant1.
\end{equation}

The Euler-Lagrange equation for \eqref{eq:energyintegrated} is then
\begin{equation}
\label{eq:eulerlagrange}
\begin{split}
 &\frac{1}{2\sigma}[2(1+\eta^{2}\sigma^{2})^{2}+4\eta^{2}\sigma^{2}-\eta^{2}\sigma^{2}\left(2+k_{3}\right)(5+\eta^{2}\sigma^{2})]a(\sigma)\\
 &\qquad= \sigma(1-\eta^{2}\sigma^{2})^2a_{,\sigma\sigma}(\sigma) + (1-\eta^{2}\sigma^{2})a_{,\sigma}(\sigma)\,,
\end{split}
\end{equation}
and, for $0\leqslant\eta<1$, the associated Robin's condition for $\sigma=1$ is given by
\begin{equation}
\label{eq:eulerlagrangeboundary}
 (1-\eta^{2})a_{,\sigma}(1)+ \left(1+\eta^{2}-2k_{24}\right)a(1)=0\,.
\end{equation}
Unfortunately, equation \eqref{eq:eulerlagrange} is too complicated to lend itself to an analytic solution, and so a rigorous study of the sign of the second variation $\mathcal{E}$ in \eqref{eq:energysigma} could not be performed, and a complete linear-stability analysis of the axial configuration within a torus remains elusive. We shall be contended with exploiting \eqref{eq:energysigma} to explore the borders of linear instability. To this end, we use a simple consequence of \eqref{eq:eulerlagrange} that must be valid for all solutions that do not vanish identically. Taking the limit as $\sigma\to0^+$, we obtain the following asymptotic form of \eqref{eq:eulerlagrange},
\begin{equation}\label{eq:euler_lagrange_asymptotic}
\frac{a(\sigma)}{\sigma}\approx\sigma a_{,\sigma\sigma}+a_{,\sigma},\quad\sigma\approx0\,.
\end{equation}
This is a homogeneous equation with solutions $a=A\sigma$ and $a=A\sigma^{-1}$, with $A$ an arbitrary constant, only the former of which  conforms with \eqref{eq:a_0=0}. Thus, the extremals of $\mathcal{E}$ that can make it negative destabilizing the axial configuration are such that 
\begin{equation}\label{eq:a=A_sigma}
a(\sigma)\approx A\sigma,\quad\text{for}\quad\sigma\approx0,\quad\text{with}\quad A\neq0\,.
\end{equation}
This condition, together with \eqref{eq:eulerlagrangeboundary},  will lead us in the following section to construct a test function that improves the linear instability analysis for toroidal nematics known from \cite{koning:saddle-splay}.

\section{Instability criteria}
 \label{sec:criteria}
Before probing in a torus the stability of the universal solution in which all integral director lines are circle, we remark that since the functional in $\mathcal{E}$ in \eqref{eq:energyintegrated} is quadratic any perturbation $a$ that succeeds in making $\mathcal{E}$ negative is accompanied by $-a$, which assigns the same value to $\mathcal{E}$. Though mathematically this is nearly a trivial remark, physically it says that when the axial configuration becomes unstable in a torus it gives way to two equally energetic (and so, equally likely) distortions that however differ in the sense their integral lines are wound around the torus. Such a \emph{chirality degeneracy}, which was discussed at length in \cite{koning:saddle-splay}, is an intrinsic feature of our stability problem. In the following, we shall only concentrate on one variant of two possible destabilizing modes, the one for which $A>0$ in  \eqref{eq:a=A_sigma}, but we should not forget that it is accompanied by its opposite twin. 

Moreover, if $\mathcal{E}[a^\ast]<0$, it follows immediately from $\mathcal{E}$ being quadratic that it can be made unbounded from below by just multiplying $a^\ast$ by an increasing constant. A normalization condition must be added to our search for admissible test functions $a^\ast$: we shall take it to be\footnote{Often, the integral of $|a^\ast|^2$ is prescribed as a normalization condition. Here, we chose to prescribe instead the derivative of $a^\ast$ at a point; the special value in \eqref{eq:a_normalization} was taken to represent the ratio of the only two lengths present in the problem.}
\begin{equation}\label{eq:a_normalization}
a^\ast_{,\sigma}(0)=\eta\,.
\end{equation} 

\begin{figure}[h!]
	\begin{center}
		\begin{subfigure}{.32\textwidth}
			\centering
			\includegraphics[width=\linewidth]{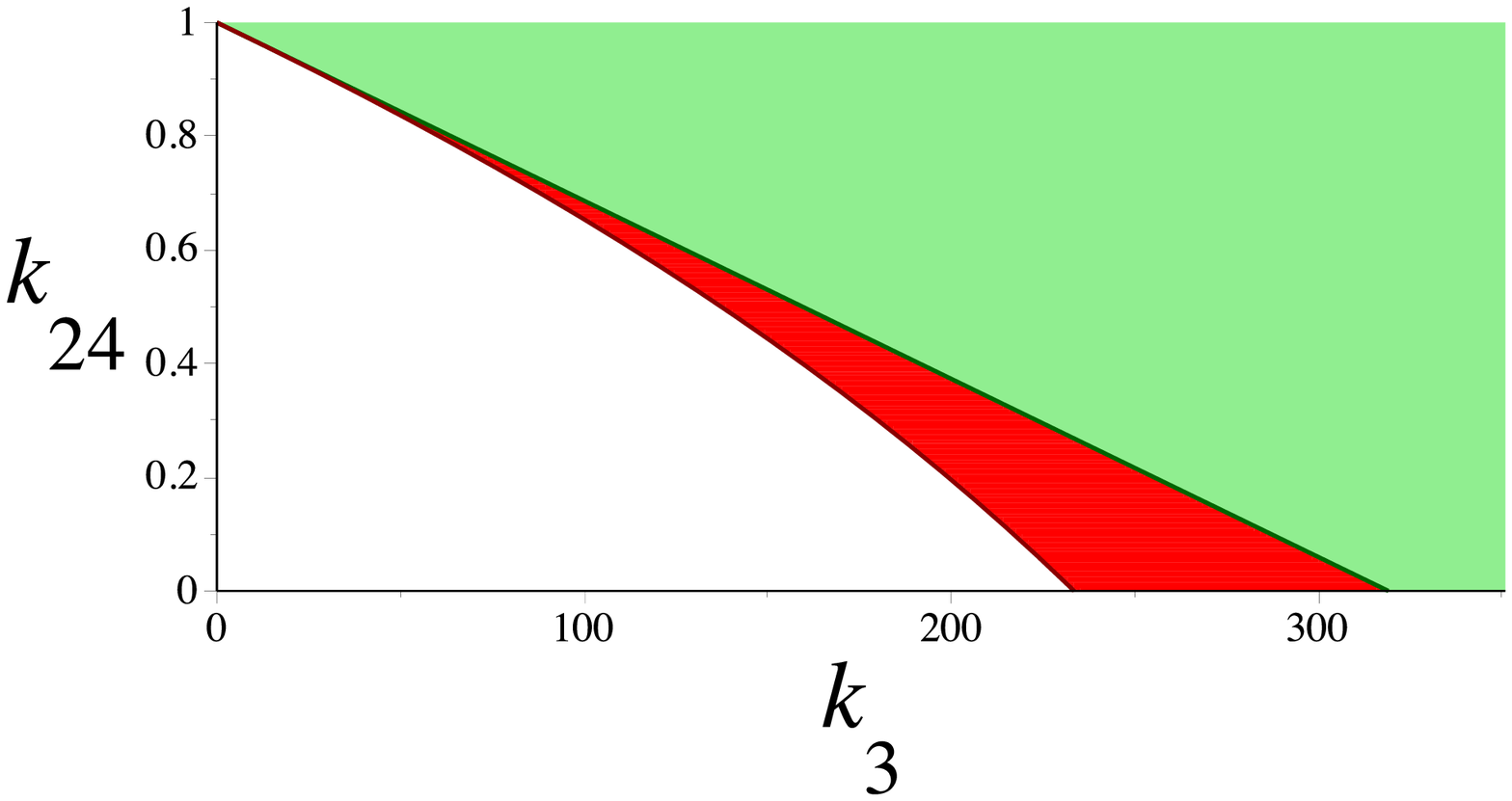}
			\caption{$\eta=0.1$}
		\end{subfigure}
		\begin{subfigure}{.32\textwidth}
			\centering
			\includegraphics[width=\linewidth]{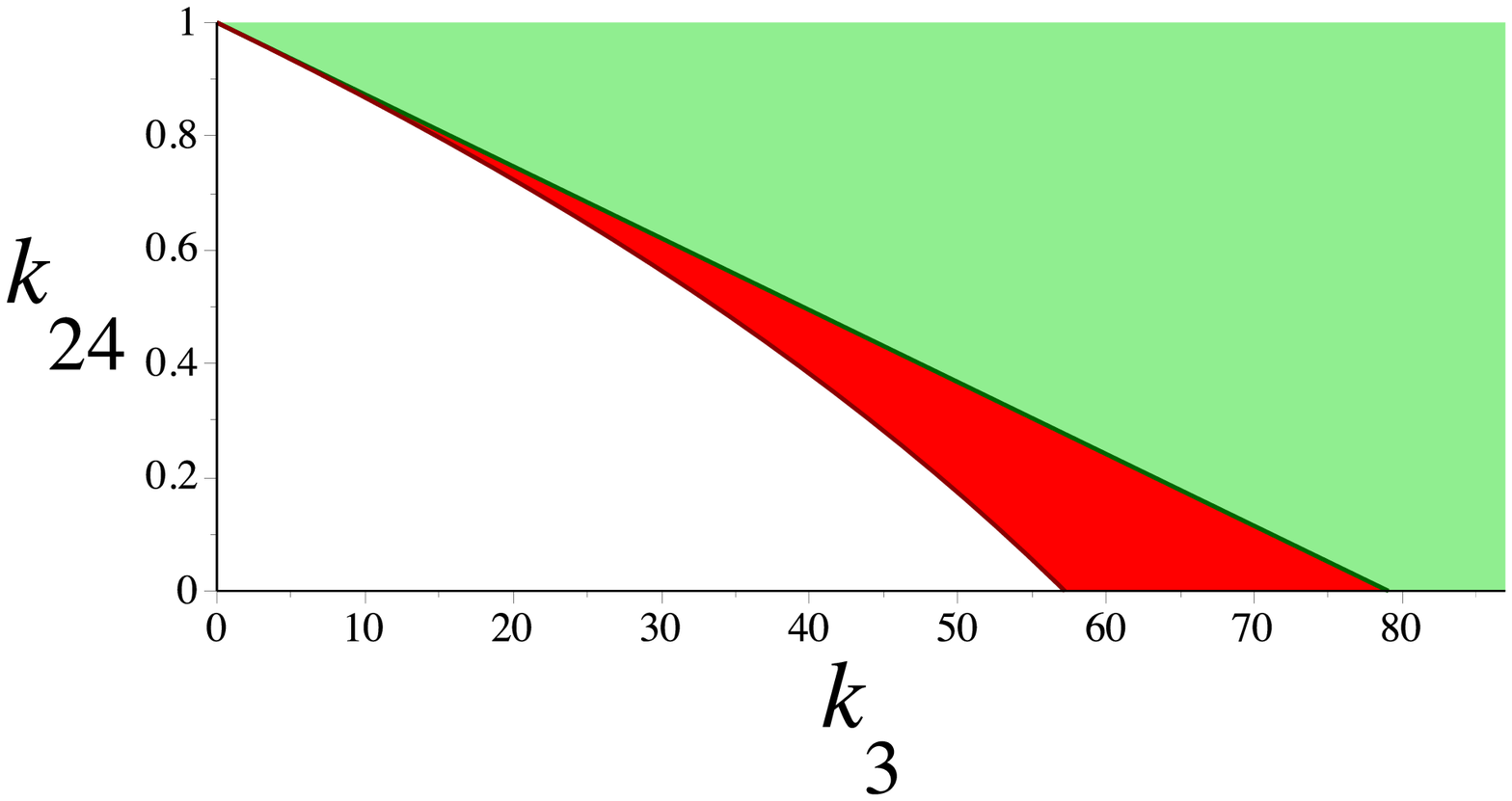}
			\caption{$\eta=0.2$}
		\end{subfigure}
		\begin{subfigure}{.32\textwidth}
			\centering
			\includegraphics[width=\linewidth]{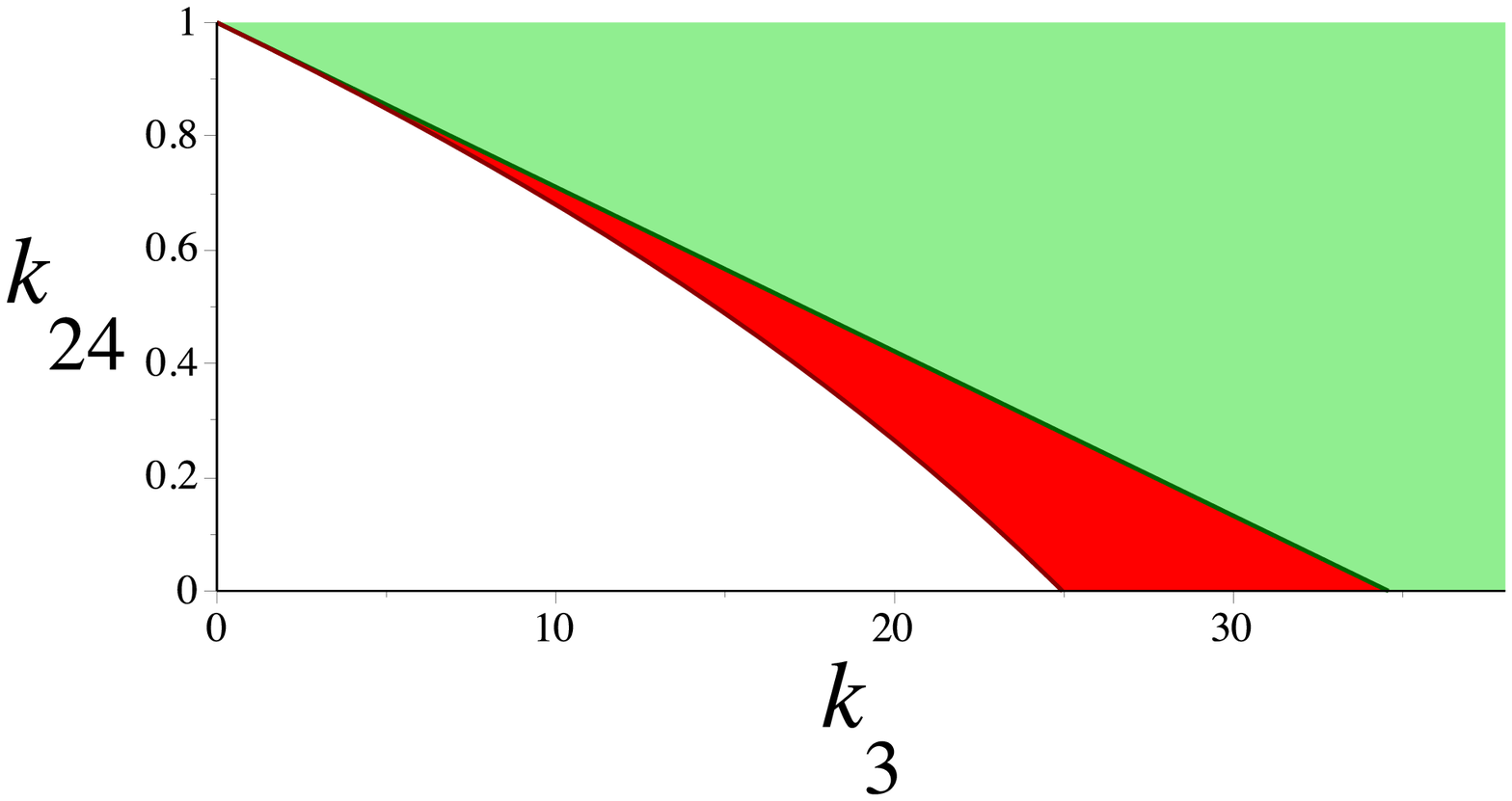}
			\caption{$\eta=0.3$}
		\end{subfigure}
		\begin{subfigure}{.32\textwidth}
			\centering
			\includegraphics[width=\linewidth]{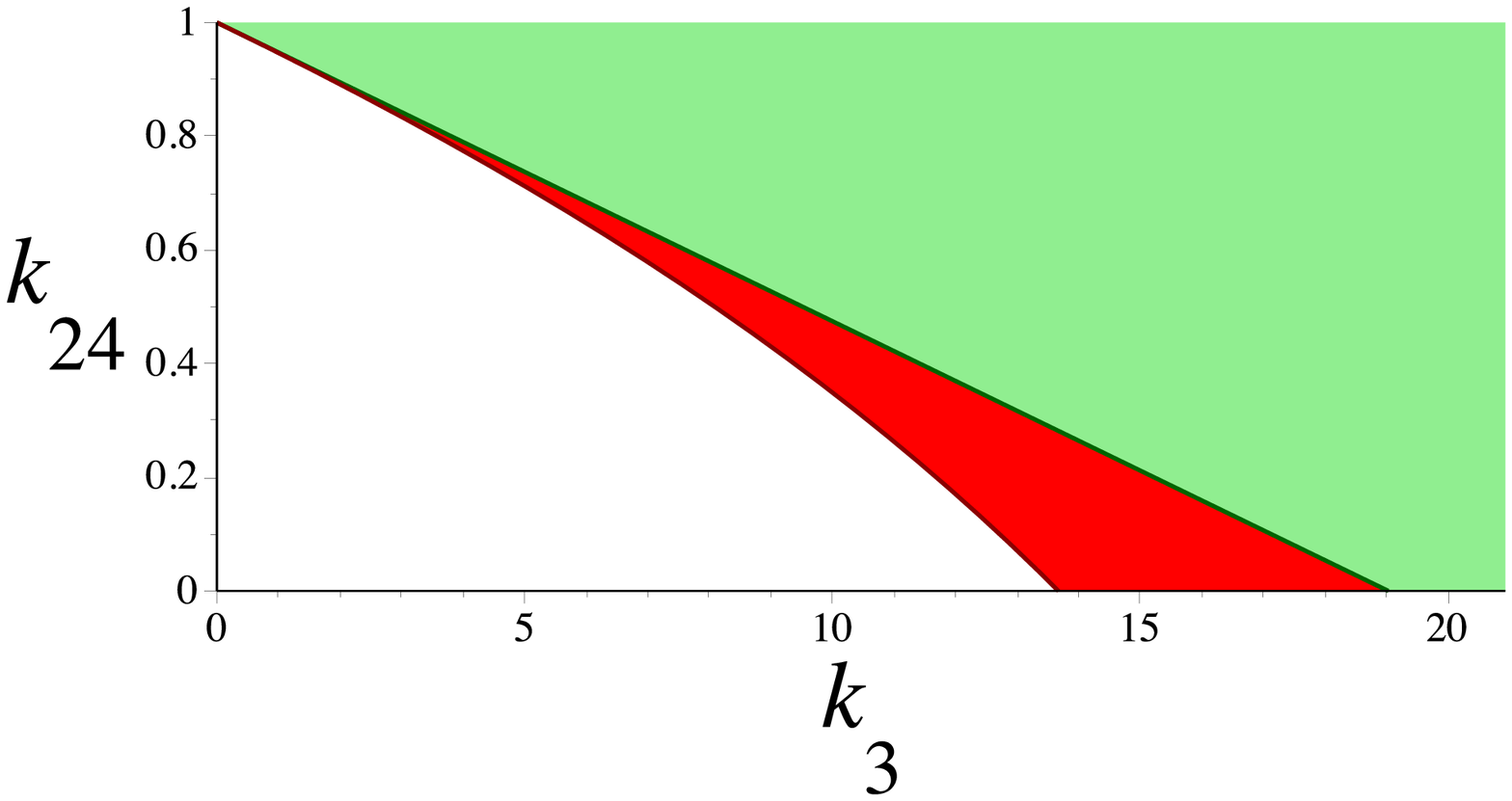}
			\caption{$\eta=0.4$}
		\end{subfigure}
		\begin{subfigure}{.32\textwidth}
			\centering
			\includegraphics[width=\linewidth]{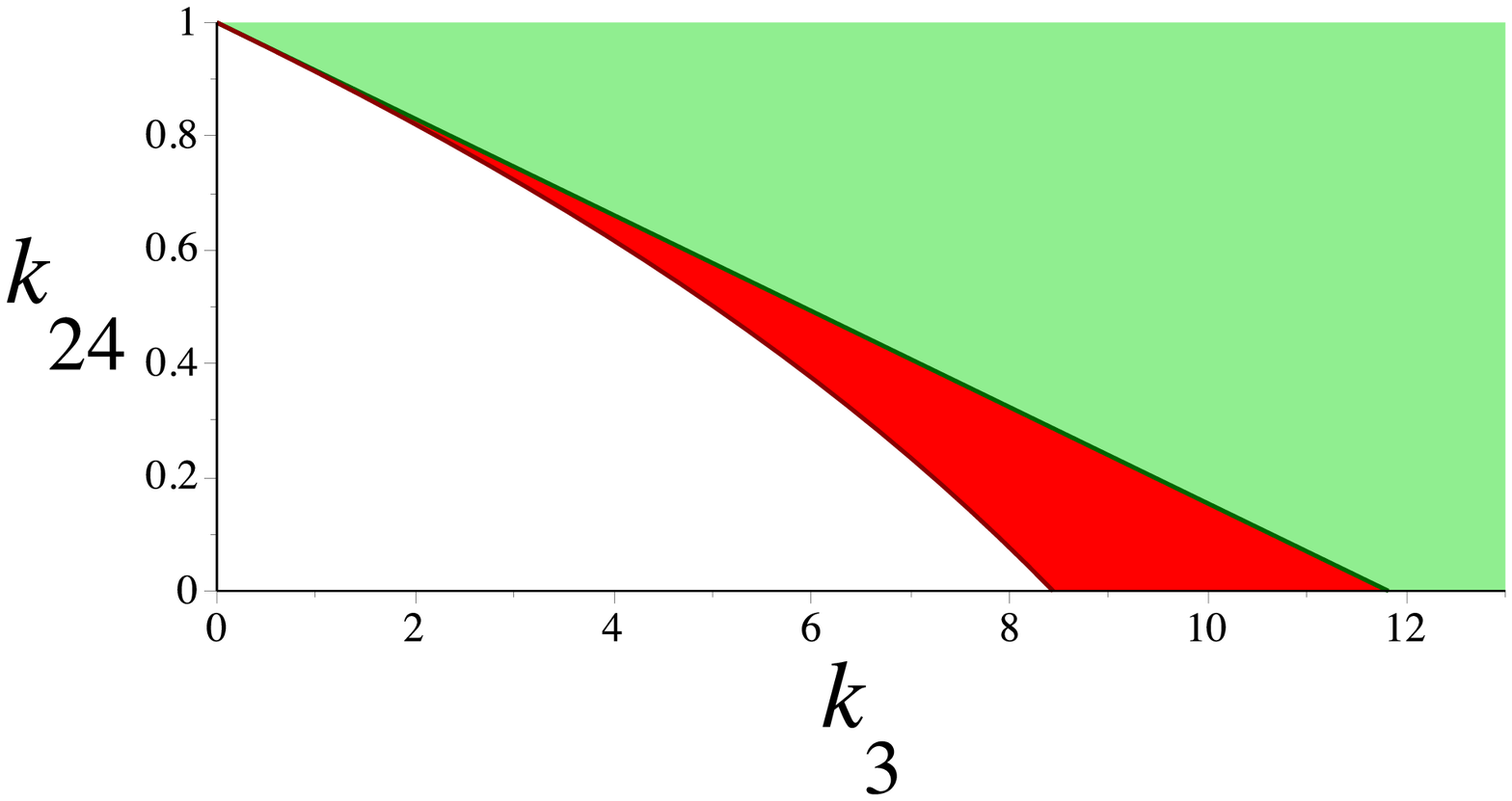}
			\caption{$\eta=0.5$}
		\end{subfigure}
		\begin{subfigure}{.32\textwidth}
			\centering
			\includegraphics[width=\linewidth]{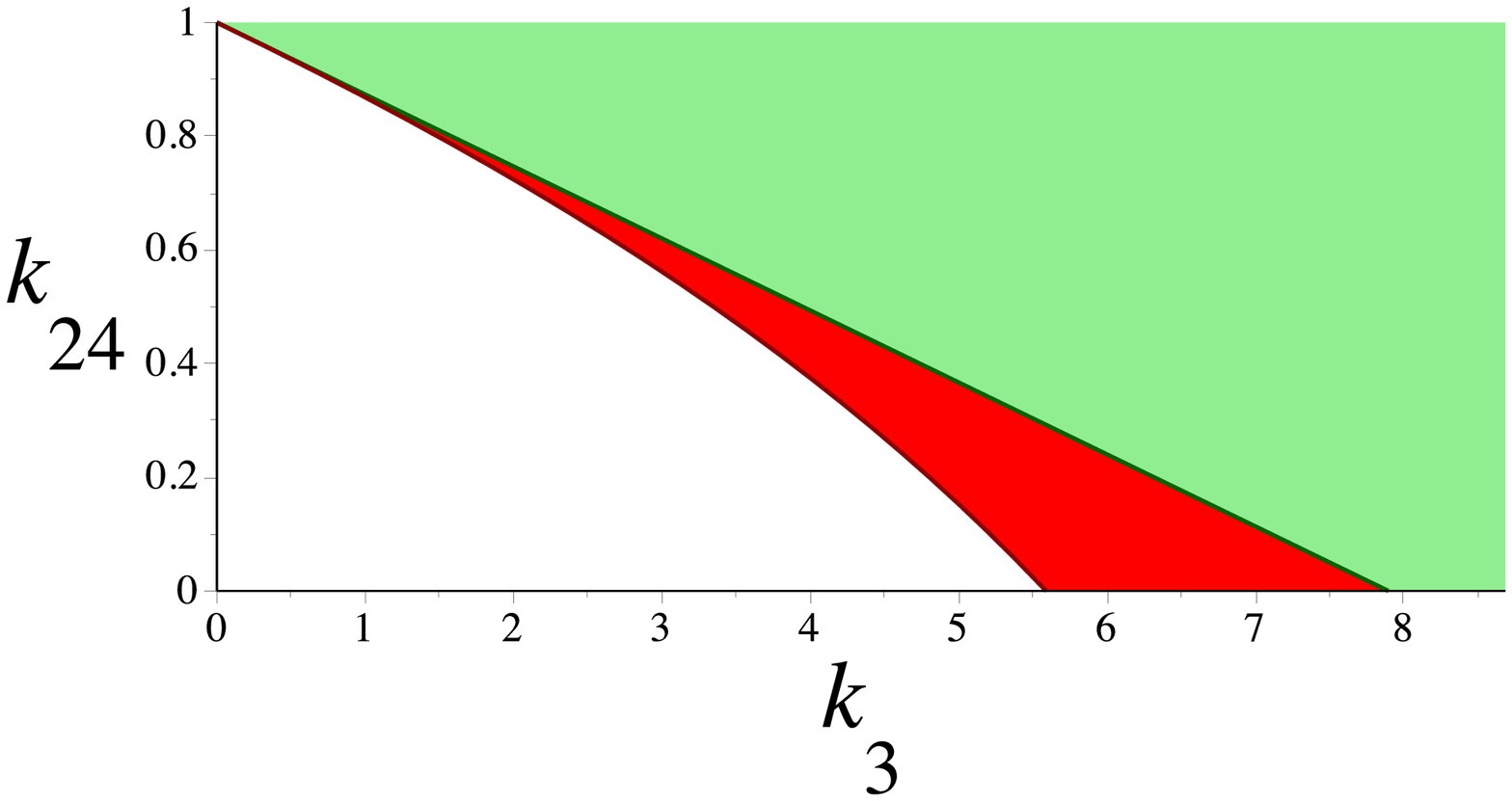}
			\caption{$\eta=0.6$}
		\end{subfigure}
		\begin{subfigure}{.32\textwidth}
			\centering
			\includegraphics[width=\linewidth]{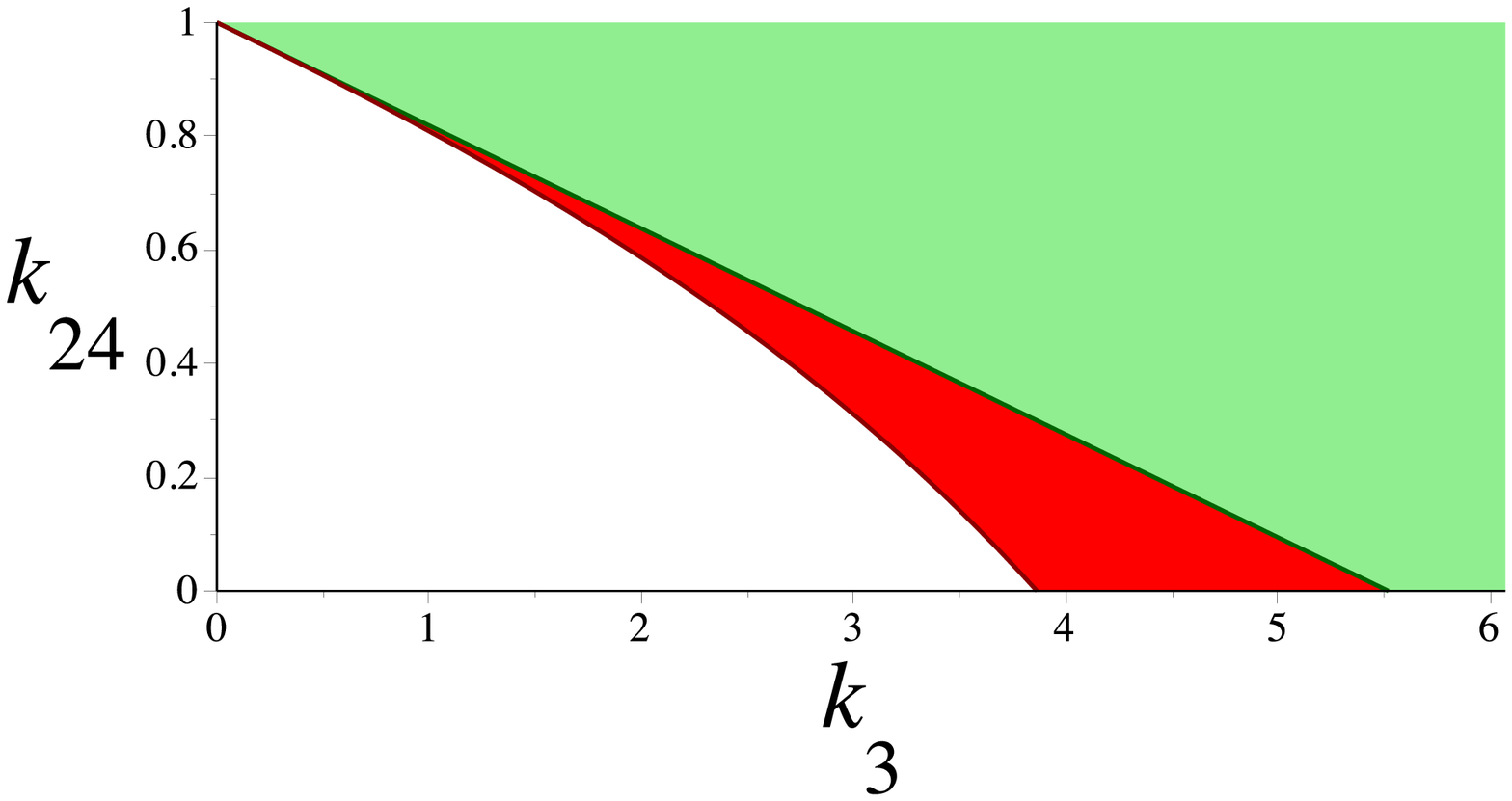}
			\caption{$\eta=0.7$}
		\end{subfigure}
		\begin{subfigure}{.32\textwidth}
			\centering
			\includegraphics[width=\linewidth]{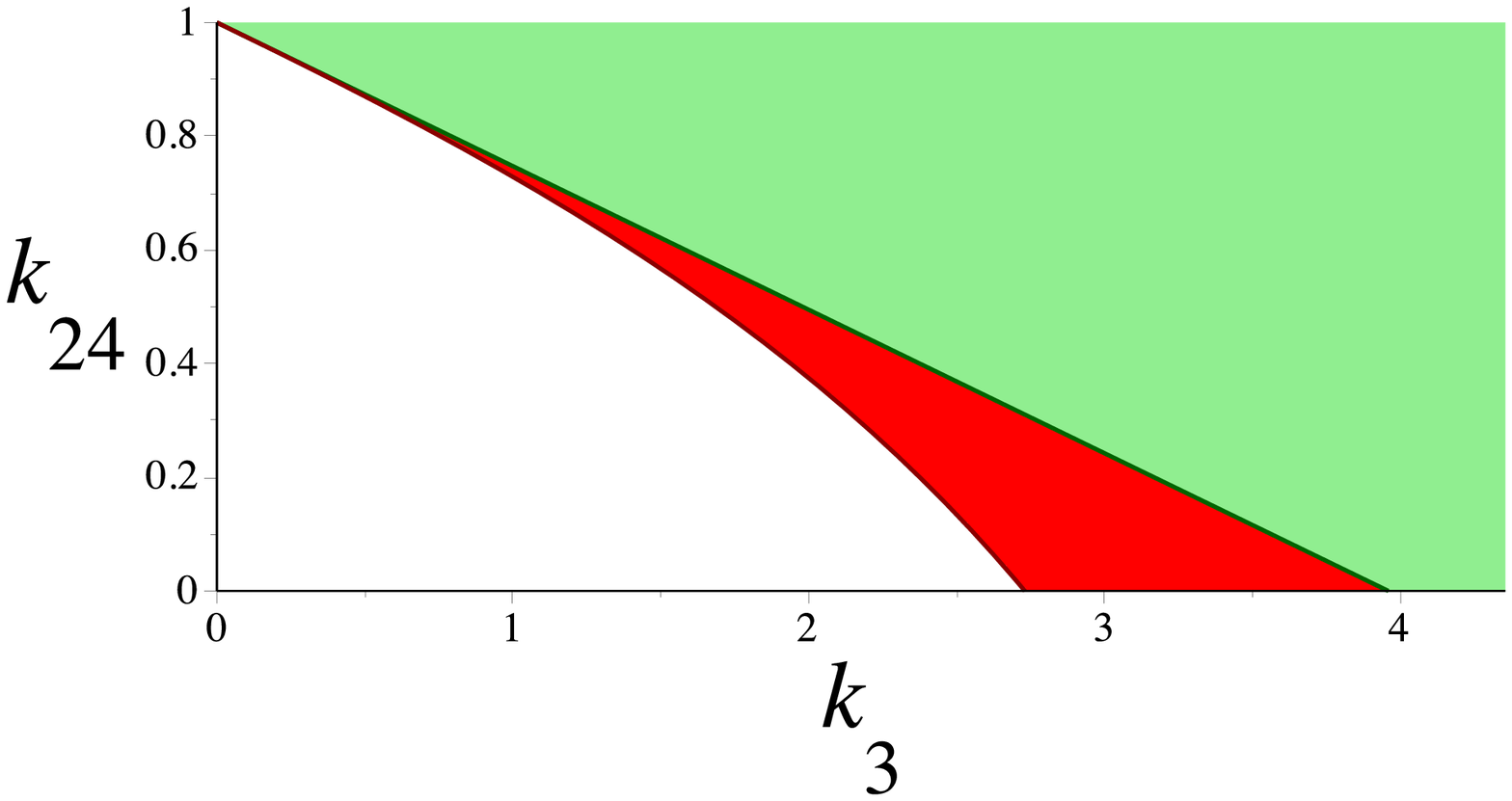}
			\caption{$\eta=0.8$}
		\end{subfigure}
		\begin{subfigure}{.32\textwidth}
			\centering
			\includegraphics[width=\linewidth]{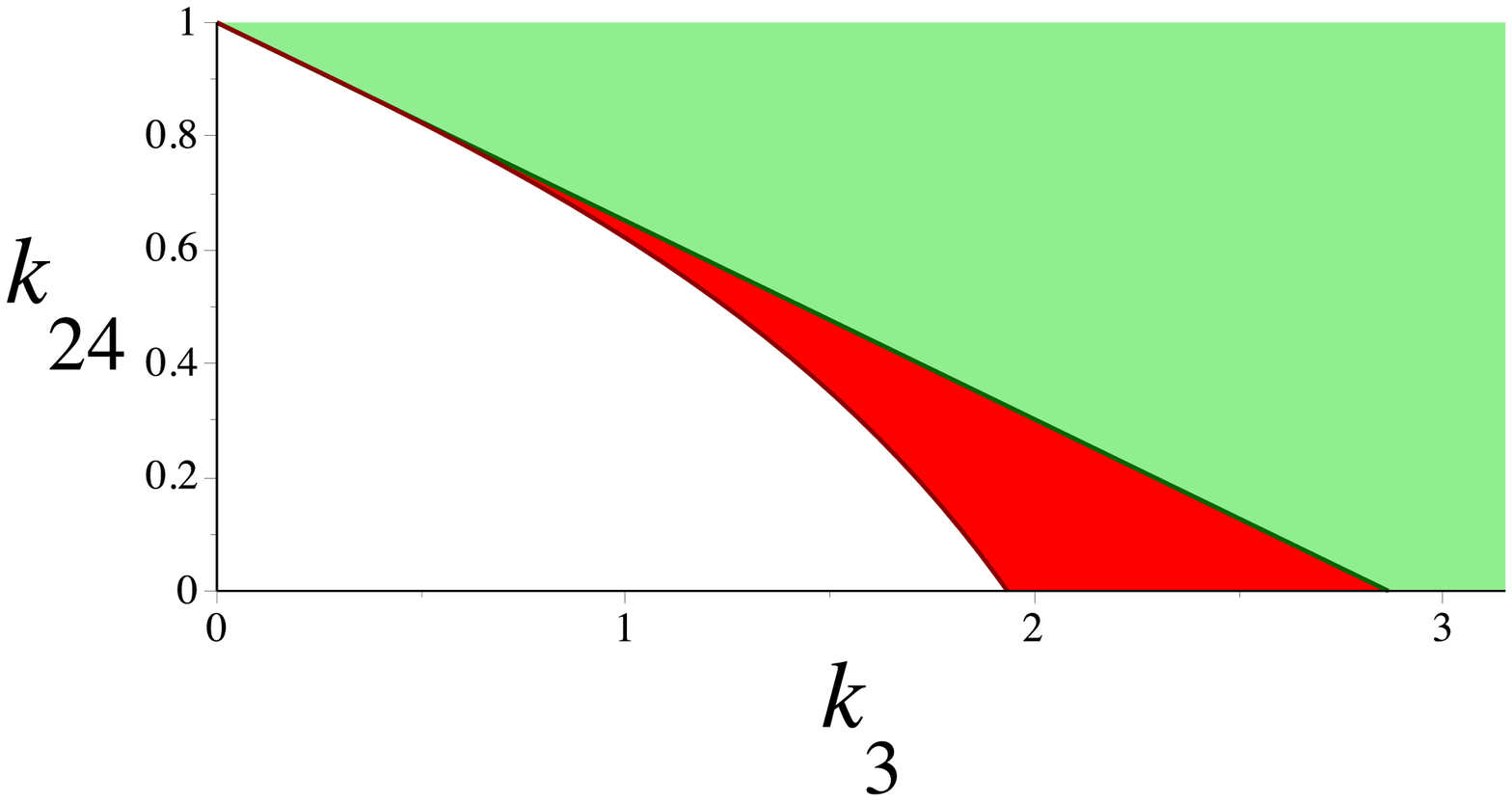}
			\caption{$\eta=0.9$}
		\end{subfigure}
		\caption{Regions of instability identified by the modes $\alin$ and $\abar$. The green region (lighter grey, in print) represents inequality \eqref{eq:koning}; the red spike (darker grey, in print) must be added to the green region to represent the more convoluted, but less restrictive inequality \eqref{eq:a_quadratic_inequality_proper}. }
			\label{fig:comparison}
	\end{center}
\end{figure}

In the light of this normalization, we may rewrite the \emph{linear} test function chosen in \cite{koning:saddle-splay} as $a^\ast=\alin:=\eta\sigma$. By requiring $\mathcal{E}[\alin]<0$, we easily arrive at the inequality 
\begin{equation}\label{eq:koning}
 k_{24} > 1 + \frac{\eta^{4}-9\eta^{2}+6-6(1-\eta^{2})^{\frac{3}{2}}}{4\eta^{2}}k_{3}\,,
\end{equation}
which, for any given $\eta$, represents the region above a straight line (with negative slope) in the strip $0\leq k_{24}\leq1$ of the plane $(k_3,k_{24})$ (see Fig.~\ref{fig:comparison}). By use of \eqref{eq:scaled_elastic_constants}, we readily reduce \eqref{eq:koning} to the form
\begin{equation}\label{eq:koning_explicit}
\frac{K_2-K_{24}}{K_3}<k_\mathrm{c}(\eta)
\end{equation}
with $k_\mathrm{c}$ precisely given by (23) of \cite{koning:saddle-splay} (after having set $\xi:=1/\eta$). The interesting physical interpretation of \eqref{eq:koning_explicit} is that the saddle-splay constant $K_{24}$ \emph{screens} the energy cost of twist, and doing so it facilitates the instability of the axial configuration, as it makes it happen for larger values of $K_2$ compared to the case $K_{24}=0$ (or strong anchoring is enforced on $\boundary$). As attractive as this interpretation may be, it geometrically relies on being the limit of stability in \eqref{eq:koning} a straight line in the plane $(k_3,k_{24})$ passing through the point $(0,1)$.

\begin{figure}[h]
	\centering
	\includegraphics[width=.7\linewidth]{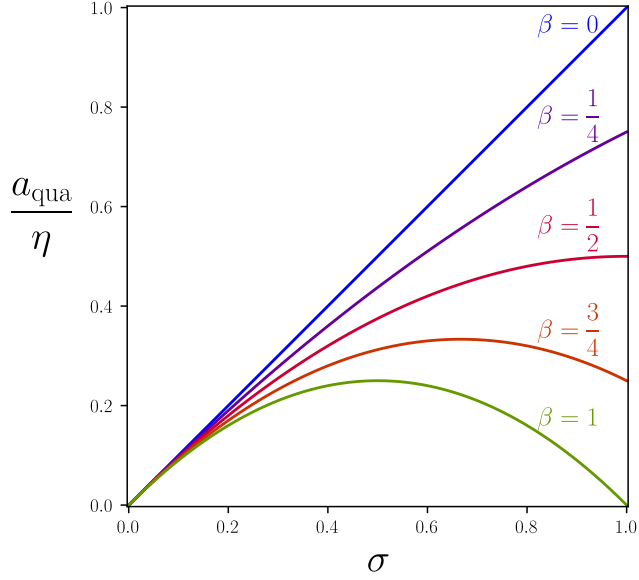}
	\caption{Graph of the function $\abar$ in \eqref{eq:a_quadratic} shown for $\beta=0,\frac14,\frac12,\frac34,1$ (with increasing degree of bending). For all admissible values of $\beta$ delivered by \eqref{eq:beta}, the graph of $\abar$ is encapsulated between the two boundary graphs shown here.}
	\label{fig:graphs}
\end{figure}

The linear test function $\alin$ cannot possibly comply with Robin's condition \eqref{eq:eulerlagrangeboundary}, which is a necessary condition for an extremum of $\mathcal{E}$. The simplest way to make it valid for a test function $a^\ast$ (together with \eqref{eq:a=A_sigma} and \eqref{eq:a_normalization}) is to choose $a^\ast$ as the \emph{quadratic} function
\begin{equation}\label{eq:a_quadratic}
\abar(\sigma):=\eta\sigma(1-\beta\sigma)\,,
\end{equation}
where $\beta$ is a constant that \eqref{eq:eulerlagrangeboundary} determines as being
\begin{equation}\label{eq:beta}
\beta=2\frac{1-k_{24}}{3-\eta^2-2k_{24}}\,.
\end{equation}
It is a simple matter to check that letting $\eta$ vary in $[0,1)$ and $k_{24}$ in $[0,1]$ makes $\beta$ cover the interval $[0,1)$. Correspondingly, the function $\abar$ exhibits the graphs shown in Fig.~\ref{fig:graphs}. 
In particular, for $\frac12<\beta\leq1$, $\abar$ has its maximum for $0<\sigma<1$. 

With some labour, we reduce the inequality $\mathcal{E}[\abar]<0$ to
\begin{subequations}\label{eq:a_quadratic_inequality}
\begin{equation}\label{eq:a_quadratic_inequality_proper}	
k_3>\frac{K(\eta,k_{24})}{H(\eta,k_{24})}\,,
\end{equation}
where
\begin{equation}\label{eq:K_function_definition}
K(\eta,k_{24}):=
4(1-k_{24})
\left[2\eta^{2}(\eta^{2}+1)k_{24}+4\left(1-\sqrt{1-\eta^{2}}\right)\left(1-k_{24}\right)+\eta^{2}\left(3\eta^{4}-5\eta^{2}-2\right)\right]\,,
\end{equation}
\begin{equation}\label{eq:H_function_definition}
\begin{split}
H(\eta,k_{24})&:=-4\left(1-k_{24}\right)^{2}
\left[\eta^{4}+55\eta^{2} +18\eta^{2}\sqrt{1-\eta^{2}} -56\left(1-\sqrt {1-\eta^{2}}\right)\right]\\
&-18(1-k_{24})\sqrt {1-\eta^{2}}\left[5 \left(2k_{24}+\eta^{2}-3 \right)\eta\arcsin \eta+4\eta^{2}\left(1-\eta^{2} \right)\right]\\
&-6\left(1-k_{24}\right)\eta^{2}(3+\eta^{2})(1-\eta^{2})+3\eta^{2}(\eta^{4}-9\eta^{2}+6)\left(1-\eta^{2}\right)
-18\eta^{2}\left( 1-\eta^{2} \right)^{\frac{5}{2}}
\end{split}
\end{equation}
\end{subequations}
Inequality \eqref{eq:a_quadratic_inequality_proper} is admittedly far less elegant than \eqref{eq:koning}, but it also improves upon that substantially for all values of $0<\eta<1$, as  the region of the plane $(k_3,k_{24})$ delimited within the admissible strip $0\leq k_{24}\leq1$ by \eqref{eq:a_quadratic_inequality} includes that delimited by \eqref{eq:koning}, as shown in the graphs of Fig.~\ref{fig:comparison}. 
By differentiating with respect to $k_{24}$ the functions on the right side of \eqref{eq:a_quadratic_inequality_proper} and \eqref{eq:koning}, we see that for all $0<\eta<1$ the graphs shown in Fig.~\ref{fig:comparison} are tangent to one another where they intersect. Therefore, of the two destabilizing modes, $\alin$ and $\abar$, the latter is finer than the former, uniformly in the geometric parameter $\eta$. 

Since the line that marks the limit of stability for $\abar$ in the plane $(k_3,k_{24})$ is \emph{not} a straight line, the neat interpretation afforded by \eqref{eq:koning_explicit} is clearly in jeopardy. However, it remains true that, for given $\eta$ and $K_2$, an increase in $K_{24}$ widens the instability domain, that is, it reduces the critical value of $K_3$ that must be exceeded to make the cost of bend larger compared to the competing cost of twist. In this region, we may say that, as was perhaps expected, the instability of pure bend is enhanced by the surface energy in \eqref{eq:franktaylor} (which, for $K_{24}>0,$ is negative). However, we cannot say that surface energy is the only drive behind this instability, because for 
\begin{equation}\label{eq:beta_greater_1o2}
k_{24}<\frac12(1+\eta^2)\,,
\end{equation}
which according to \eqref{eq:beta} makes $\beta>\frac12$, the destabilizing mode $\abar$ exhibits its maximum twist in the interior of the torus, not on its boundary.
\section{Conclusions}
 \label{sec:conclusion}
In a broad sense, this paper is a contribution to the linear stability of universal equilibrium solutions for Frank's elastic energy of nematic liquid crystals. We probed the stability of the axial configuration within a circular torus with planar degenerate anchoring conditions on its boundary. We found that the domain of instability for this solution is indeed broader than shown in a previous study \cite{koning:saddle-splay}.

This is not the only outcome of our study. Not only does the new destabilizing mode $\abar$ in \eqref{eq:a_quadratic} broaden systematically the instability domain detected in \cite{koning:saddle-splay}, but it also shows a new, unexpected qualitative character of the instability. In the parameter range identified by \eqref{eq:beta_greater_1o2}, the maximum twist of the destabilizing mode is not achieved on the boundary of the torus, which would assign to the surface-like elastic constant $K_{24}$ the role of main drive of the destabilization, but it occurs well inside the torus, which exalts the role of the bulk-like elastic constants $K_2$ and $K_3$. This new scenario, which our linear analysis unveiled for the germ of instability, is likely to herald a property of the minimizers of the full-blown variational problem, where the energy functional has the form in \eqref{eq:frankalpha}. The analysis of this highly non-linear problem is presently underway, guided by the outcomes of the linear stability analysis presented here.

Having shown that the surface-like elastic constant $K_{24}$ is not the main driving force behind the instability of the bend-rich axial configuration in a torus poses a new question. Can elastic anisotropy, and more precisely a sufficiently small value of the ratio of the twist to bend constants, $K_2/K_3$, be alone responsible for the instability of the universal axial configuration in a torus with strong anchoring conditions on its boundary? This question, which we propend to answer for the positive, will be addressed in a subsequent paper, in a fully non-linear setting.

The mathematical character of our study should not prevent the reader from appreciating its potential applications to novel materials. In most thermotropic nematic liquid crystals, Frank's elastic constants $K_1$, $K_2$, and $K_3$ are nearly equal in value \cite[p.\,105]{degennes:physics}. However, in a wide class of lyotropic liquid crystals, called \emph{chromonics}, which have only recently been discovered and fully characterized \cite{zhou:elasticity}, $K_2$ may be as small as one tenth the (almost common) value of $K_1$ and $K_3$. This explains, at least heuristically, a number of recent puzzling experiments, which have shown an unexpected excess of twist deformation arising in the  equilibrium textures \cite{davidson:chiral,davidson:erratum,jeong:chiral_structures,jeong:chiral_symmetry,lubensky:confined} (see also the witty review \cite{masters:chromonic}).
Now,  both excess of twist and smallness of $K_2$ are the leitmotifs of our paper. This makes chromonics the natural experimental test-bed for our analytical results.  

 \section*{Acknowledgements}
 The work of A.P. has been supported by the University of Pavia under the PRG initiative, meant to foster research among young postdoctoral fellows. E.G.V. acknowledges the kind hospitality of the Oxford Centre for Nonlinear PDE, where part of this work was done while he was visiting the Mathematical Institute at the University of Oxford.  We are both grateful to A. Zarnescu for several enlightening discussions on the subject of this paper while our project was in its early stages.
\appendix
\section{Toroidal geometry and energy}
  \label{app:geometry}
In this Appendix, we collect a number of technical details  used in 	in Sects.~\ref{sec:toroidal} and \ref{sec:variation}, which are omitted there for ease of presentation.

\subsection{Construction and properties of the toroidal frame}
 \label{app:geometrylocalbasis}
Let $(\bm{e}_{r},\bm{e}_{\varphi},\bm{e}_{\psi})$ be the toroidal frame corresponding to the coordinates $\torcoo$ defined in \eqref{eq:positionvector}. Letting $\torcoo$ depend on a parameter $t$ makes $\bm{p}$ describe a trajectory in three-dimensional space. Differentiating this with respect to $t$, we obtain
\begin{equation*}
\begin{split}
 \dot{\bm{p}} &= \dot{r}\, (\cos\varphi\cos\psi\,\bm{e}_{x}+\sin\varphi\cos\psi\,\bm{e}_{y}+\sin\psi\,\bm{e}_{z})\\
 &\quad+ (R_{1}+r\cos\psi) \dot{\varphi}\, (-\sin\varphi\,\bm{e}_{x} + \cos\varphi\,\bm{e}_{y})\\
 &\quad+ r \dot{\psi}\, (-\cos\varphi\sin\psi\,\bm{e}_{x}-\sin\varphi\sin\psi\,\bm{e}_{y}+\cos\psi\,\bm{e}_{z})\,,
\end{split}
\end{equation*}
from which, setting $\dot{\bm{p}}=q_{r}\e_r+q_{\varphi}\e_\varphi+q_{\psi}\e_\psi$, we extract both
the toroidal frame $\torframe$ as defined in \eqref{eq:localbasis}, and $q_{r} = \dot{r}$, $q_{\varphi} = (R_{1}+r\cos\psi) \dot{\varphi}$, $q_{\psi} = r \dot{\psi}$, which readily deliver \eqref{eq:area_volume_elements}. It is immediate to check that \eqref{eq:localbasis} provides an orthonormal and positively oriented basis of the Euclidean space $\R^{3}$. Moreover,
\begin{equation}
\label{eq:localbasisderivatives}
 \left\{\!
 \begin{aligned}
  \dot{\bm{e}}_{r} &=  \cos\psi\dot{\varphi}\,\bm{e}_{\varphi} + \dot{\psi}\,\bm{e}_{\psi}\,,\\
  \dot{\bm{e}}_{\varphi} &= -\cos\psi\dot{\varphi}\,\bm{e}_{r} + \sin\psi\dot{\varphi}\,\bm{e}_{\psi}\,,\\
  \dot{\bm{e}}_{\psi} &= -\dot{\psi}\,\bm{e}_{r} - \sin\psi\dot{\varphi}\,\bm{e}_{\varphi}\,.
 \end{aligned}
 \right.
\end{equation}

We recall that for any scalar field $\chi(\bm{p})$ expressed in the toroidal frame $(\bm{e}_{r},\bm{e}_{\varphi},\bm{e}_{\psi})$:
\begin{equation*}
\dot{\chi}= \grad\chi\cdot\dot{\bm{p}} = \chi_{,r}\dot{r} + \chi_{,\varphi}\dot{\varphi} + \chi_{,\psi}\dot{\psi}\,.
\end{equation*}
Then $\dot{\bm{e}}_{r}=\grad\bm{e}_{r}\,\dot{\bm{p}}$, $\dot{\bm{e}}_{\varphi}=\grad\bm{e}_{\varphi}\,\dot{\bm{p}}$, and $\dot{\bm{e}}_{\psi}=\grad\bm{e}_{\psi}\,\dot{\bm{p}}$ and, since 
$\dot{\bm{p}} = \dot{r}\,\bm{e}_{r} + (R_{1}+r\cos\psi)\dot{\varphi}\,\bm{e}_{\varphi} + r\dot{\psi}\,\bm{e}_{\psi}$, it is immediate to check that
\begin{equation}
\label{eq:localbasisgradients}
 \left\{\!
 \begin{aligned}
 \grad\bm{e}_{r} &= \frac{\cos\psi}{R_{1}+r\cos\psi}\bm{e}_{\varphi}\otimes\bm{e}_{\varphi} + \frac{1}{r}\bm{e}_{\psi}\otimes\bm{e}_{\psi}\,, \\
 \grad\bm{e}_{\varphi} &= -\frac{\cos\psi}{R_{1}+r\cos\psi}\bm{e}_{r}\otimes\bm{e}_{\varphi} +\frac{\sin\psi}{R_{1}+r\cos\psi}\bm{e}_{\psi}\otimes\bm{e}_{\varphi}\,, \\
 \grad\bm{e}_{\varphi} &= -\frac{\sin\psi}{R_{1}+r\cos\psi}\bm{e}_{\varphi}\otimes\bm{e}_{\varphi} -\frac{1}{r}\bm{e}_{r}\otimes\bm{e}_{\psi}\,. 
\end{aligned}
\right.
\end{equation}
\subsection{Gradient, divergence, and curl of the director field}
 \label{app:geometrydirectorfield}
We consider a director nematic field $\n$ expressed in the toroidal frame as in~\eqref{eq:director}.
Recalling that for any vector field $\bm{\chi}(\bm{p}) \das \chi_{r}\,\bm{e}_{r} + \chi_{\varphi}\,\bm{e}_{\varphi} + \chi_{\psi}\,\bm{e}_{\psi}$
\begin{equation*}
 \grad\bm{\chi} = \chi_{r}\grad\bm{e}_{r} + \bm{e}_{r}\otimes\grad \chi_{r} + \chi_{\varphi}\grad\bm{e}_{\varphi} + \bm{e}_{\varphi}\otimes\grad \chi_{\varphi} +\chi_{\psi}\grad\bm{e}_{\psi} + \bm{e}_{\psi}\otimes\grad \chi_{\psi}\,,
\end{equation*}
\eqref{eq:localbasisgradients} yields 
\begin{equation*}
\begin{split}
 \grad\bm{n} &= n_{r,r}\,\bm{e}_{r}\otimes\bm{e}_{r} + \frac{n_{r,\varphi} - \cos\psi\,n_{\varphi}}{R_{1}+r\cos\psi}\,\bm{e}_{r}\otimes\bm{e}_{\varphi} + \frac{n_{r,\psi}-n_{\psi}}{r}\,\bm{e}_{r}\otimes\bm{e}_{\psi} \\
 &\quad+ n_{\varphi,r}\,\bm{e}_{\varphi}\otimes\bm{e}_{r} + \frac{\cos\psi\,n_{r}+n_{\varphi,\varphi}-\sin\psi\,n_{\psi}}{R_{1}+r\cos\psi}\,\bm{e}_{\varphi}\otimes\bm{e}_{\varphi} + \frac{n_{\varphi,\psi}}{r}\,\bm{e}_{\varphi}\otimes\bm{e}_{\psi} \\
 &\quad+ n_{\psi,r}\,\bm{e}_{\psi}\otimes\bm{e}_{r} + \frac{\sin\psi\,n_{\varphi}+n_{\psi,\varphi}}{R_{1}+r\cos\psi}\,\bm{e}_{\psi}\otimes\bm{e}_{\varphi} + \frac{n_{r}+n_{\psi,\psi}}{r}\,\bm{e}_{\psi}\otimes\bm{e}_{\psi}\,.
\end{split}
\end{equation*}
Therefore
\begin{equation}\label{eq:appendix_div_n}
\begin{split}
 \dv\bm{n} &= n_{r,r} + \frac{R_{1} + 2r\cos\psi}{r(R_{1}+r\cos\psi)}\,n_{r} + \frac{1}{R_{1}+r\cos\psi}\,n_{\varphi,\varphi} \\
 &\quad - \frac{\sin\psi}{R_{1}+r\cos\psi}n_{\psi}+ \frac{1}{r}n_{\psi,\psi}
\end{split}
\end{equation}
and
\begin{equation*} 
\begin{split}
 \curl\bm{n} &= \left(\frac{\sin\psi\,n_{\varphi}+n_{\psi,\varphi}}{R_{1}+r\cos\psi} - \frac{n_{\varphi,\psi}}{r}\right)\bm{e}_{r} + \left(\frac{n_{r,\psi}-n_{\psi}}{r} - n_{\psi,r}\right)\bm{e}_{\varphi} \\
 &\quad + \left(n_{\varphi,r} - \frac{n_{r,\varphi} - \cos\psi\,n_{\varphi}}{R_{1}+r\cos\psi}\right)\bm{e}_{\psi}\,.
\end{split}
\end{equation*}
Thus
\begin{equation}\label{eq:appendix_abnormality_n}
\begin{split}
 \bm{n}\cdot\curl\bm{n} &= \frac{\sin\psi\,n_{r}\,n_{\varphi} + n_{r}\,n_{\psi,\varphi} - n_{\psi}\,n_{r,\varphi}}{R_{1}+r\cos\psi} - \frac{R_{1}}{r(R_{1}+r\cos\psi)}n_{\varphi}n_{\psi} \\
 &\quad+ \frac{n_{\varphi}\,n_{r,\psi} - n_{r}\,n_{\varphi,\psi}}{r} - n_{\varphi}n_{\psi,r} + n_{\psi}n_{\varphi,r}\,,
\end{split}
\end{equation}
\begin{equation*}
\begin{split}
-\n\times\curl\n=(\grad\bm{n})\bm{n} &= \left(n_{r}\,n_{r,r} + \frac{n_{\varphi}\,n_{r,\varphi} - \cos\psi\,n_{\varphi}^{2}}{R_{1}+r\cos\psi} + \frac{n_{\psi}\,n_{r,\psi} - n_{\psi}^{2}}{r}\right)\bm{e}_{r} \\
 &\quad+ \left(n_{r}\,n_{\varphi,r} + \frac{\cos\psi\,n_{r}\,n_{\varphi} - \sin\psi\,n_{\varphi}\,n_{\psi} + n_{\varphi}\,n_{\varphi,\varphi}}{R_{1}+r\cos\psi} + \frac{n_{\psi}\,n_{\varphi,\psi}}{r}\right)\bm{e}_{\varphi} \\
 &\quad+ \left(n_{r}\,n_{\psi,r} + \frac{\sin\psi\,n_{\varphi}^{2} + n_{\varphi}\,n_{\psi,\varphi}}{R_{1}+r\cos\psi}  + \frac{n_{r}\,n_{\psi} + n_{\psi}\,n_{\psi,\psi}}{r}\right)\bm{e}_{\psi}\,.
\end{split}
\end{equation*}
\subsection{Frank's energy}
 \label{app:geometryfrankenergy}
For brevity, we shall write $\Free = \dfrac{1}{2}\left(\Free_{1} + \Free_{2} + \Free_{3}\right) + \Free_{24}$, where 
\begin{gather*}
\Free_{1}[\n] \das K_{1}\int_{\body}(\dv\bm{n})^{2}\dd\volume\,, \quad \Free_{2}[\n] \das K_{2}\int_{\body}(\bm{n}\cdot\curl\bm{n})^{2}\dd\volume\,,\\
\Free_{3}[\n] \das K_{3}\int_{\body}|(\grad\bm{n})\bm{n}|^{2}\dd\volume \quad\text{and}\quad \Free_{24}[\n] \das K_{24}\int_{\boundary}[(\grads\bm{n})\bm{n}-(\dvs\bm{n})\bm{n}]\cdot\bm{\nu}\dd\area\,.
\end{gather*}
We now compute $\Free$ for the director field $\n$ described in Sect.~\ref{app:geometrydirectorfield} relative to  the toroidal frame $\torframe$, under the simplifying assumption that  $\vartheta=\frac{\pi}{2}$ (i.e.\ when $n_{r} = 0$, $n_{\varphi} = \cos\alpha$ and $n_{\psi} = \sin\alpha$) and $\alpha=\alpha(r,\psi)$. We obtain
\begin{subequations}\label{F_i_defintions}
\begin{equation}\label{eq:F_1_definition}
 \mathcal{F}_{1}[\alpha] = K_{1} \int_{0}^{R_{2}}\!\int_{0}^{2\pi}\!\int_{0}^{2\pi}\frac{\left[r\sin\psi\sin\alpha - (R_{1}+r\cos\psi)\cos\alpha\,\alpha_{,\psi}\right]^{2}}{r(R_{1}+r\cos\psi)}\dd r\dd\varphi\dd\psi\,,
\end{equation}
\begin{equation}\label{eq:F_2_definition}
 \mathcal{F}_{2}[\alpha] = K_{2} \int_{0}^{R_{2}}\!\int_{0}^{2\pi}\!\int_{0}^{2\pi}\frac{\left[r(R_{1}+r\cos\psi)\alpha_{,r} + R_{1}\sin\alpha\cos\alpha\right]^{2}}{r(R_{1}+r\cos\psi)}\dd r\dd\varphi \dd\psi\,,
\end{equation}
\begin{equation}\label{eq:F_3_definition}
\begin{split}
 \mathcal{F}_{3}[\alpha] = K_{3} \int_{0}^{R_{2}}\!\int_{0}^{2\pi}\!\int_{0}^{2\pi}&\left\{\frac{\left[r\sin\psi\cos\alpha + (R_{1}+r\cos\psi)\sin\alpha\,\alpha_{,\psi}\right]^{2}}{r(R_{1}+r\cos\psi)}\right. \\
 &\qquad \left.+ \frac{\left[r\cos\psi + R_{1}\sin^{2}\alpha\right]^{2}}{r(R_{1}+r\cos\psi)}\right\}\dd r\dd\varphi \dd\psi\,,
\end{split}
\end{equation}
\begin{equation}\label{eq:F_24_definition}
 \mathcal{F}_{24}[\alpha] = - K_{24} \int_{0}^{2\pi}\!\int_{0}^{2\pi}\left(R_{2}\cos\psi + R_{1}\sin^{2}\alpha(R_{2},\psi)\right)\dd\varphi\dd\psi\,.
\end{equation}
\end{subequations}
By using \eqref{eq:eta_definition} and \eqref{eq:sigma_definition},
and taking $\alpha=\alpha(\sigma,\psi)$, we get $\dd r=R_{1}\eta\dd\sigma$ and $\alpha_{,r}=\frac{1}{R_{1}\eta}\alpha_{,\sigma}$. Then, by computing the integral in $\varphi$, the four components of Frank's energy become
\begin{equation*}
 \mathcal{F}_{1}[\alpha] = 2\pi R_{1}K_{1}\eta \int_{0}^{1}\!\int_{0}^{2\pi}\frac{\left[\eta\sigma\sin\psi\sin\alpha - (1+\eta\sigma\cos\psi)\cos\alpha\,\alpha_{,\psi}\right]^{2}}{\eta\sigma(1+\eta\sigma\cos\psi)}\dd\sigma d\psi\,,
\end{equation*}
\begin{equation*}
 \mathcal{F}_{2}[\alpha] = 2\pi R_{1}K_{2}\eta \int_{0}^{1}\!\int_{0}^{2\pi}\frac{\left[\sigma(1+\eta\sigma\cos\psi)\alpha_{,\sigma} + \sin\alpha\cos\alpha\right]^{2}}{\eta\sigma(1+\eta\sigma\cos\psi)}\dd\sigma d\psi\,,
\end{equation*}
\begin{equation*}
\begin{split}
 \mathcal{F}_{3}[\alpha] = 2\pi R_{1}K_{3}\eta \int_{0}^{1}\!\int_{0}^{2\pi}&\left\{\frac{\left[\eta\sigma\sin\psi\cos\alpha + (1+\eta\sigma\cos\psi)\sin\alpha\,\alpha_{,\psi}\right]^{2}}{\eta\sigma(1+\eta\sigma\cos\psi)}\right. \\
 &\qquad \left.+ \frac{\left[\eta\sigma\cos\psi + \sin^{2}\alpha\right]^{2}}{\eta\sigma(1+\eta\sigma\cos\psi)}\right\}\dd\sigma d\psi\,,
\end{split}
\end{equation*}
\begin{equation*}
 \mathcal{F}_{24}[\alpha] = - 2\pi R_{1}K_{24} \int_{0}^{2\pi}\sin^{2}\alpha(1,\psi)\dd\psi\,,
\end{equation*}
and \eqref{eq:frankalpha} follows at once. Letting $\alpha\equiv0$ in \eqref{eq:frankalpha}, we readily arrive at
\begin{equation}\label{eq:F_0_computed}
\free[0]=\pi R_1K_3\int_{0}^{1}\!\int_{0}^{2\pi}\frac{\eta^{2}\sigma}{1+\eta\sigma\cos\psi}\dd\sigma \dd\psi  = 2\pi^2R_1K_3(1-\sqrt{1-\eta^2})\,,
\end{equation}
valid for all $0\leqslant\eta\leqslant1$.
\subsection{Quadratic approximation}
 \label{app:geometrytaylor}
Letting  $f_{1}$, $f_{2}$, $f_{3}$, and $f_{24}$ denote  the integrands  in the functionals $\mathcal{F}_{1}$, $\mathcal{F}_{2}$, $\mathcal{F}_{3}$, and $\mathcal{F}_{24}$, respectively, see \eqref{F_i_defintions}, we approximate them to the second order in $\alpha$ as
\begin{gather*}
 f_{1} \approx \frac{\eta\sigma\sin^{2}\psi}{1+\eta\sigma\cos\psi}\alpha^{2} + \frac{1+\eta\sigma\cos\psi}{\eta\sigma}\alpha_{,\psi}^{2} - 2\sin\psi\,\alpha\alpha_{,\psi}\,,\\
 f_{2} \approx \frac{1}{\eta\sigma(1+\eta\sigma\cos\psi)}\alpha^{2} + \frac{\sigma(1+\eta\sigma\cos\psi)}{\eta}\alpha_{,\sigma}^{2} +\frac{2}{\eta}\alpha\alpha_{,\sigma}\,,\\
 f_{3} \approx \frac{\eta\sigma}{1+\eta\sigma\cos\psi} + \frac{2\cos\psi - \eta\sigma\sin^{2}\psi}{1+\eta\sigma\cos\psi}\alpha^{2} + 2\sin\psi\,\alpha\alpha_{,\psi}\,,\\
 f_{24} \approx \alpha^{2}(1,\psi)\,,
\end{gather*}
where both \eqref{eq:eta_definition} and \eqref{eq:sigma_definition} have been used and a factor $R_1$ has been pulled out of all integrals.
Since $2\sin\psi\,\alpha\alpha_{,\psi} = \sin\psi\,(\alpha^2)_{,\psi}$, an integration by parts shows that
\begin{equation*}
 \int_{0}^{2\pi}2\sin\psi\,\alpha\alpha_{,\psi}\dd\psi = - \int_{0}^{2\pi}\cos\psi\,\alpha^2\dd\psi\,
\end{equation*}
and so the bending energy can be given the simple form in \eqref{eq:franktaylor}.

To derive \eqref{eq:energyintegrated} from \eqref{eq:energysigma}, a number of integrals in $\psi$ must be computed; they are recorded here for completeness, so as to allow the interested reader to retrace all our steps:
\begin{align*}
\int_{0}^{2\pi}\frac{(1-\eta\sigma\cos\psi)^{2}}{(1+\eta\sigma\cos\psi)^{3}}\dd\psi &= 2\pi\frac{(1+\eta^{2}\sigma^{2})^{2} + 2\eta^{2}\sigma^{2}}{(1-\eta^{2}\sigma^{2})^{\frac{5}{2}}}\,,\\
\int_{0}^{2\pi}\frac{1-\eta\sigma\cos\psi}{(1+\eta\sigma\cos\psi)^{2}}\dd\psi &= 2\pi\frac{1+\eta^{2}\sigma^{2}}{(1-\eta^{2}\sigma^{2})^{\frac{3}{2}}}\,,\\
\int_{0}^{2\pi}\frac{1}{1+\eta\sigma\cos\psi}\dd\psi &= 2\pi\frac{1}{(1-\eta^{2}\sigma^{2})^{\frac{1}{2}}}\,,\\
\int_{0}^{2\pi}\frac{\cos\psi-\eta\sigma}{(1+\eta\sigma\cos\psi)^{3}}\dd\psi &= -\pi\frac{(5+\eta^{2}\sigma^{2})\eta\sigma}{(1-\eta^{2}\sigma^{2})^{\frac{5}{2}}}\,,\\
\int_{0}^{2\pi}\frac{1}{(1+\eta\sigma\cos\psi)^{2}}\dd\psi &= 2\pi\frac{1}{(1-\eta^{2}\sigma^{2})^{\frac{3}{2}}}\,.
\end{align*}
Similarly, it is advisable to perform the following integration by parts:
\begin{equation*}
\int_{0}^{1}2\frac{1+\eta^{2}\sigma^{2}}{(1-\eta^{2}\sigma^{2})^{\frac{3}{2}}}a(\sigma)a_{,\sigma}(\sigma)\dd\sigma  
=  \frac{1+\eta^{2}}{(1-\eta^{2})^{\frac{3}{2}}}a^{2}(1) - \int_{0}^{1}\frac{\eta^{2}\sigma(5+\eta^{2}\sigma^{2})}{(1-\eta^{2}\sigma^{2})^{\frac{5}{2}}}a^{2}(\sigma)\dd\sigma\,,
\end{equation*}
where use has also been made of \eqref{eq:a_0=0}.

\end{document}